\pdfoutput=1

\documentclass[11pt]{article}

\usepackage[final]{ACL2023}

\usepackage{times}
\usepackage{latexsym}

\usepackage[T1]{fontenc}

\usepackage[utf8]{inputenc}

\usepackage{microtype}

\usepackage{inconsolata}

\usepackage{inconsolata}
\usepackage[utf8]{inputenc} 
\usepackage[T1]{fontenc}    
\usepackage{hyperref}       
\usepackage{url}            
\usepackage{booktabs}       
\usepackage{amsfonts}       
\usepackage{nicefrac}       
\usepackage{microtype}      
\usepackage{xcolor}         

\usepackage{tikz}
\usepackage{lipsum}
\usepackage{float}
\usepackage{multirow}
\usepackage{color}
\usepackage{hhline}
\usepackage{algpseudocode}
\usepackage{algorithm}
\usepackage{setspace}
\usepackage[most]{tcolorbox}
\usepackage{float}
\usepackage{colortbl}
\usepackage{cancel}
\usepackage{soul}
\usepackage{amsmath}
\usepackage{bm}
\usepackage{amssymb}
\usepackage{youngtab}
\newcommand\scalemath[2]{\scalebox{#1}{\mbox{\ensuremath{\displaystyle #2}}}}
\usepackage {caption}
\usepackage{multibib}
\usepackage{dblfloatfix}
\usepackage{float}

\newcommand\encircle[1]{%
  \tikz[baseline=(X.base)] 
    \node (X) [draw, shape=circle, inner sep=0, fill=black, text=white] {\strut #1};%
}

\usepackage{times}
\usepackage{latexsym}
\usepackage[T1]{fontenc}
\usepackage[utf8]{inputenc}
\usepackage{microtype}

\title{Threat Behavior Textual Search by Attention Graph Isomorphism}

 \author{Chanwoo Bae,\quad Guanhong Tao,\quad Zhuo Zhang,\quad Xiangyu Zhang \\
         Purdue University, USA \\
         \{bae68, taog, zhan3299\}@purdue.edu, xyzhang@cs.purdue.edu}

\begin{document}
\maketitle

\maketitle
\begin{abstract}
Cyber attacks cause over \$1 trillion loss every year. 
An important task for cyber security analysts is attack forensics. It entails understanding malware behaviors and attack origins. However, existing automated or manual malware analysis can only disclose a subset of behaviors due to inherent difficulties (e.g., malware cloaking and obfuscation). As such, analysts often resort to text search techniques to identify existing malware reports based on the symptoms they observe, exploiting the fact that malware samples share a lot of similarity, especially those from the same origin. In this paper, we propose a novel malware behavior search technique that is based on graph isomorphism at the attention layers of Transformer models. We also compose a large dataset collected from various agencies to facilitate such research. Our technique outperforms state-of-the-art methods, such as those based on sentence embeddings and keywords by 6-14\%.  In the case study of 10 real-world malwares, our technique can correctly attribute 8 of them to their ground truth origins while using Google only works for 3 cases.

\end{abstract}

\section{Introduction}
Cyber-attacks are a prominent threat to our daily life, causing over \$1 trillion loss every year. Defending and mitigating cyber-attacks are hence critical. An important task in the arms race is attack forensics, which aims to determine malware behaviors, damages, and origins. It usually starts with an attack instance, e.g., a malware sample captured in the wild. The analysts use tools such as IDA~\cite{ferguson2008reverse} to inspect its code body, and sand-boxing techniques such as Cuckoo~\cite{oktavianto2013cuckoo} to execute it and observe its runtime behaviors.
Attack forensics are important because the results can be used to assess damages and prevent future attacks.
However, malware often employs sophisticated self-protection such as obfuscation~\cite{you2010malware} that changes code body to make it difficult to understand and/or masquerade a benign application, and cloaking that conceals malicious payload until certain (attack) conditions are satisfied.
As a result, analysts usually can only disclose a part of malware behaviors. They hence heavily rely on text search to find existing related malware reports. Such search is usually driven by the observed behaviors such as sabotage, data exfiltration (regarding how they are performed?). 
The rationale is that cyber-attacks become increasingly organized (e.g., sponsored at a state-level), showing a substantial level of commonality in terms of the exploits used (i.e., bugs in target systems that allow the malware to penetrate), the payloads delivered, and their objectives, especially for those launched by a same {\em threat actor}~\cite{malpedia} (i.e., an adversary or an organization of adversaries).
As such, one can predict a new malware's full behaviors from reports of existing malware samples that share some commonality with the new sample.
In fact, major security vendors have published a large volume of malware analysis reports. While they have the great potential to provide collective intelligence for future analysis, 
there has been an intrinsic barrier to fully leveraging such knowledge, namely, these threat reports are written in unstructured and informal natural languages. Since anyone can contribute such reports, it is difficult to standardize them. 

Therefore, the key problem is a specialized text search challenge, which is called {\em cyber threat intelligence} (CTI) search following the terminology used in the domain. 
CTI search poses two main challenges; (i) supervised-learning is hardly feasible due to the lack of labeled datasets, and (ii) existing pre-trained general-purpose language models cannot effectively capture domain specific semantics. Sometimes, small changes for a general-purpose language model denote substantial semantic differences in CTI. For example, a {\em "file"} may describe either information stealing ({\em "\textbf{file} to leak the stolen data from the program"}) or program exploitation ({\em "\textbf{file} to exploit the program for stealing data"}). 

The most popular search method directly uses  {\em indicators of compromise} (IoCs) of the malware sample, e.g., malware file hash~\cite{catakoglu2016automatic, liao2016acing}.
It is the method used in VirusTotal~\cite{virustotal}, a widely used malware analysis platform. 
Although using IoCs is precise and free from false positives,  it cannot deal with the well known malware mutation problem~\cite{liao2016acing} in which malware frequently and consistently changes its configurations, payloads, and even attack steps, to evade detection or simply update its functionalities.
Another method is text similarity based malware behavior search. 
Existing text similarity methods largely fall into two categories, {\em keywords based} methods~\cite{corley2005measuring, harispe2015semantic} and {\em sentence embedding based} methods~\cite{le2014distributed, lau2016empirical, devlin2018bert, reimers2019sentence}.
The former focuses on domain specific keywords. It cannot effectively extract relations across keywords, which are critical in CTI search. In the above example, the keyword {\em "file"} needs to be analyzed with the relation of other words (i.e., {\em "leak"} or {\em "exploit"}) - keywords-based search (i.e., \textit{"steal"}, \textit{"program"}, \textit{"data"}) will cause misunderstanding.
In contrast, directly using embeddings tends to be unnecessarily distracted by the words that are not critical to CTI search.

We propose a novel CTI search technique. We collect a large repository of CTI reports from multiple agencies such as Kaspersky, Symantec and MacAfee. Specifically, Mitre ATT\&CK~\cite{mitreattack} is a widely-known knowledge base of adversary techniques (i.e., behaviors) based on real-world malware observations. The repo covers reports in the past 20 years. We then use a {\em masked-language model} (MLM) based on Transformer to perform unsupervised learning on the dataset. We observe that the language model can pay special attention to IoC related words, and more importantly, their correlations. After training,
instead of using the pre-trained embeddings, which are noisy due to the large natural language vocabulary, 
we construct a {\em attention graph} in which a node is a word token and an edge is introduced between two nodes when their attention is larger than a threshold. 
We then use graph similarity to determine CTI report similarity.
We make the following contributions.
\begin{itemize}
  \itemsep0em 
  \item We collect a large volume of existing CTI reports from reputable sources which could facilitate future research. 
 \item We propose a novel attention graph based search method for CTI reports. It is particularly suitable in capturing the domain specific semantics of these reports. 
  \item We 
  compare our method with \textit{doc2vec}~\cite{le2014distributed, lau2016empirical} (a sentence embedding based technique), keyword based text similarity methods~\cite{corley2005measuring, harispe2015semantic},
  and a few state-of-the-art unsupervised learning based methods~\cite{reimers2019sentence}. 
  Our method consistently outperforms these baselines. 
  \item 
  In a case study of 10 real-world malware attacks, our search successfully finds the most relevant reports (from the past) that allow us to attribute 8 of these attacks to their true origins. 
  In contrast, using \textit{Google} can only correctly attributes 3 and a simple  IoC-based search correctly attributes 2. One of the generative LLMs, GPT-4 (Google Bing) correctly answers 3. 
\end{itemize}

\begin{figure*}[tb]
    \centering
    \vspace{3mm}
    {%
    \setlength{\fboxsep}{2pt}%
    \setlength{\fboxrule}{0.5pt}%
    \includegraphics[clip=true,trim=0mm 0mm 0mm 0mm,width=\linewidth]{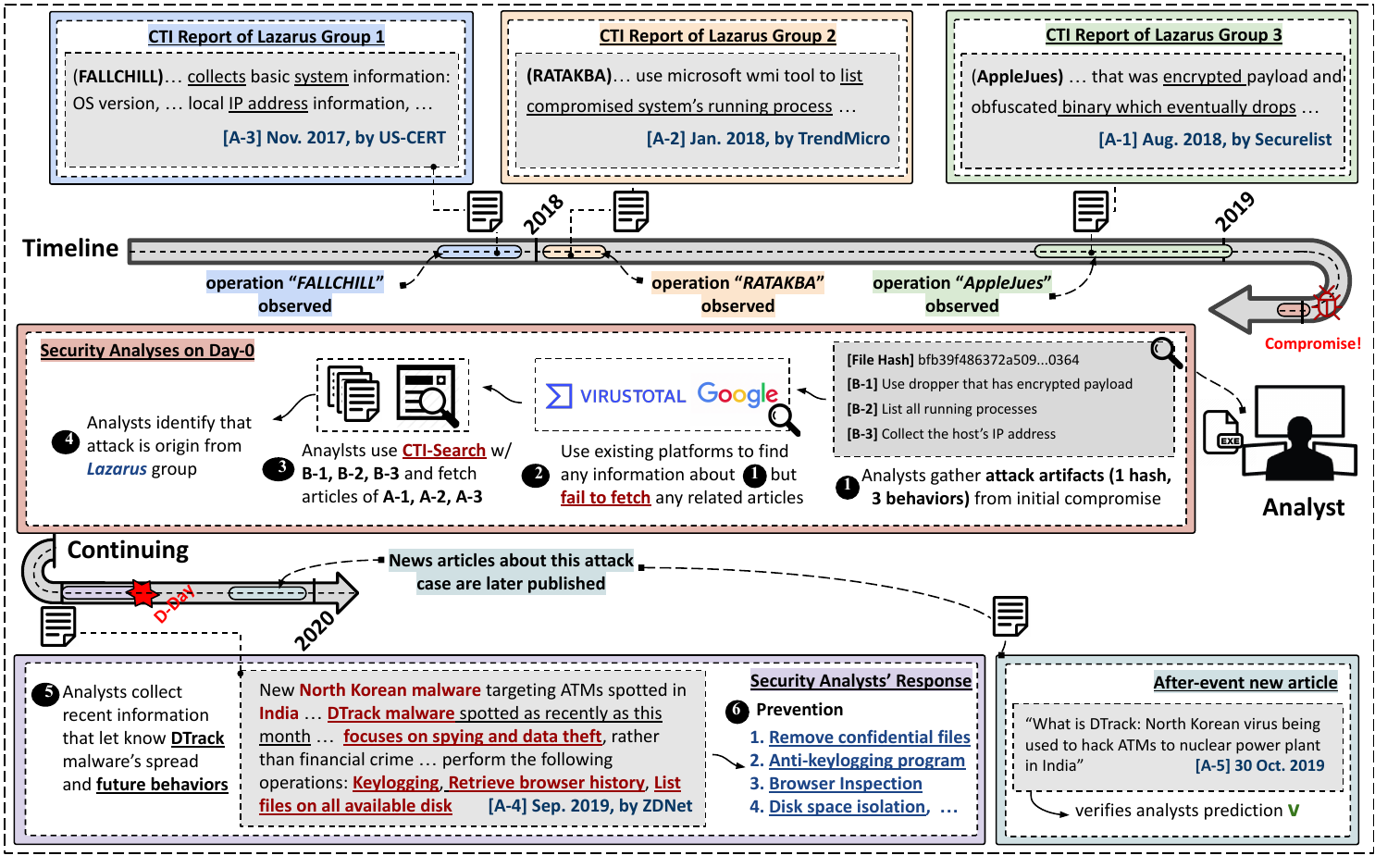}
    }%
    \vspace{-5mm}   
	\caption{Motivation example: searching a real-world attack on an Indian nuclear plant. The arrow from left to right denotes the timeline. The attack happened in 2019 (the right-most spot on the timeline with a bug symbol). A few other attacks by the same threat actor were conducted before the 2019 attack and denoted by the blue, orange and green durations along the timeline. The large box ``{\tt Security Analyses on Day-0}'' in the middle denotes the multiple methods the analyst could have used to analyze and search the attack. The boxes in the bottom show the real analysis reports of the attack there were produced long after the attack in 2020. Most of the information in those reports is covered by the {\em past} reports A-1, A-2, and A-3 retrieved by our method, illustrating that with our method, the attack could have been easily analyzed and attributed.}
	\label{fig:motivation-example-dtrack-lazarus}
\end{figure*}

\section{Motivation}

A real-world event in 2019 on a nuclear power plant in India~\cite{indiatoday2019dtrack} illustrates how an information retrieval (i.e., our system) contributes to cyberattack investigations. It is a multi-stage attack that first penetrates some computers on the power-plant's network, leveraging a zero-day code vulnerability (e.g., a bug in browser) and then laid low and silently compromise more systems leveraging normal functionalities. The process may take days or even weeks. The payload was finally delivered,  3-5 days after the initial penetration, accessing the nuclear plant's confidential data. Such complex and multi-staged attacks are also called {\em advanced persistent threat} (APT)~\cite{milajerdi2019holmes}. Assume a few days after the attack was initiated, security analysts noticed some system anomaly. Further assume they acquired an attack artifact, which is the executable malware file used in the first penetration step. From the file, analysts acquired the  hash of the malware \textrm{bfb39f486372a509...0364} and a few malware behaviors using a sandbox tool, namely, \textbf{(B-1)} \textit{"use a dropper that has encrypted payload"}, \textbf{(B-2)} \textit{"list all running processes"}, and \textbf{(B-3)} \textit{"collect the host's IP address"}. This corresponds to step \encircle{1} in Figure~\ref{fig:motivation-example-dtrack-lazarus}).

However, from these symptoms, the analysts can hardly determine the objective and scope of the attack. Since the power plant is critical infrastructure, they  need answers to a number of questions, for example, what is the attack origin (is it from a major known threat actor)? and what are the attacker's ultimate interests (e.g., infrastructure sabotage and confidential information leak)? 
Critical decisions need to be made based on the answers to these questions.  

The collected evidence is insufficient to answer these questions, which is very typical due to the inherent difficulties in malware analysis.
The analysts usually resort to CTI search In our case, assume they  first looked up the file hash on \textit{VirusTotal} to check if the same attack has been conducted in the past.
However, the attack was unique in 2019 and hence VirusTotal returned no match.
In fact, the malware was first submitted to VirusTotal on October 28, 2019, one month after the initial attack.
Then in step \encircle{2}, the analysts tried to find previous CTI reports using the behaviors, that is, B-1, B-2, and B-3. 
As the behaviors were written in natural language, they needed to rely on  text search methods.
Assume they used  Google and the textual descriptions of the behaviors.
However, the search results were not informative. 
Observe most of the retrieved items are not even semantically related. The semantically related items are in fact not related to the attack at all (refer to Appendix~\ref{section:more_on_motivation_examples}).

\textbf{Our Method.}
Assume that analysts had our search technique in 2019. 
Searching the three behaviors using our method, the analysts managed to retrieve 3 malware reports (i.e., A-1, A-2 and A-3) dated before the attack time which are all conducted by \textit{Lazarus} group within last 2 years (see top three boxes in Figure~\ref{fig:motivation-example-dtrack-lazarus}). 
The analysts hence suspected the origin of attack was the \textit{Lazarus} group (step \encircle{4}). More importantly, recent threat intelligence article (\textbf{A-4}~\cite{zdnet2019atm}) says that \textit{Lazarus} groups recently perform the following attacks: (i) browser history collection, (ii) keylogging and (iii) disk-drive scrapping (step \encircle{5}).
Therefore, the system administrator could employ the corresponding countermeasures (step \encircle{6}).

One month after the attack, real forensics reports were produced for the attack, named {\em Dtrack}~\cite{economictimes2019dtrack}.
They indicated that the attack mainly focused on {\em stealing data from the keystroke (i.e., keyboard)}, {\em monitoring web-browsing history}, {\em data dump from local disk}. The information could have been disclosed by our technique much earlier if it was available at that time. 

\begin{figure*}[tb]
     \includegraphics[clip=true,trim=0mm 0mm 0mm 0mm,width=\linewidth]{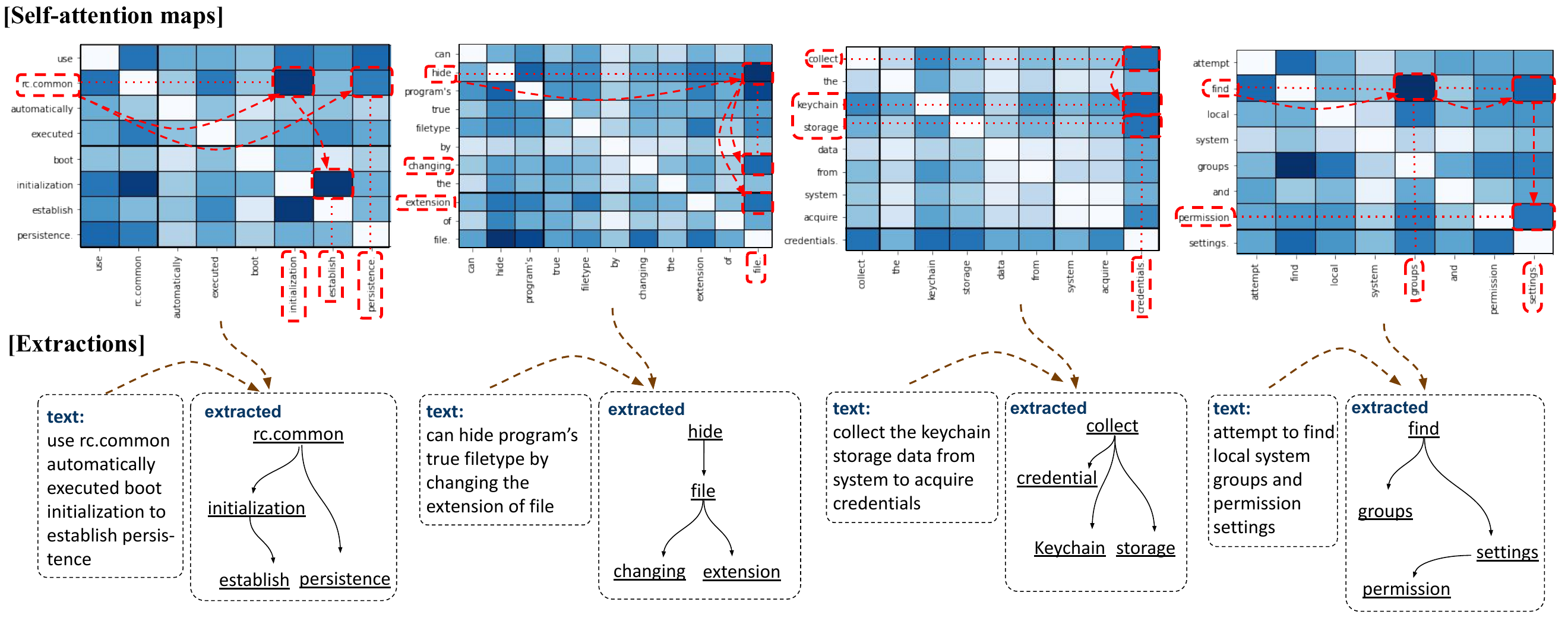} 

    \vspace{-2mm}
       
	\caption{Figures show self-attention maps on examples (top). Based on word-to-word correlations in attention map, above examples show that we can extract the core representation of behaviors (bottom) from plain text.}
	\label{fig:heat-map-example}
 	\vspace{-2mm}
\end{figure*}

\section{Related Work}
\textbf{Cyber Threat Intelligence Search.}
The most popular CTI search method uses  IoC information~\cite{liao2016acing}.  
Existing work usually formulates the challenge as  a {\em named entity recognition} (NER) problem, aiming to identify malware artifacts (e.g., \textit{IP address} and \textit{file hash}) from natural language documents~\cite{liao2016acing, zhu2018chainsmith}.
Attack ontology was proposed in~\cite{husari2017ttpdrill}, aiming to formalize malware behaviors. Researchers have proposed to  extend the data-sources of threat intelligence, such as \textit{Twitter} or \textit{Darkweb}~\cite{khandpur2017crowdsourcing, choshen2019language, wang2020into, jin2022shedding}.

\noindent
\textbf{Text  Similarity.}
Similarly analysis is the key technique behind text search, which has been well studied~\cite{corley2005measuring, islam2008semantic, budanitsky2006evaluating, mihalcea2006corpus, ramage2009random, croce2011structured, rahutomo2012semantic, kenter2015short, harispe2015semantic, rao2019bridging}.
Basically, these approaches try to capture the common keywords between two texts. Some work adopts several optimization techniques; using word-weighting~\cite{corley2005measuring, kenter2015short, lopez2019word} with IDF-score~\cite{ramos2003using}, leveraging external knowledge~\cite{islam2008semantic, budanitsky2006evaluating} and use of word similarity methods~\cite{corley2005measuring, islam2008semantic, kenter2015short, harispe2015semantic}. In the CTI search, a knowledge-based approach is limited due to absence of such resources (e.g., {\em Wordnet}~\cite{miller1995wordnet}). Meanwhile, word similarity and weighting are limited to the corpus-based approaches which are included as our baselines~\cite{corley2005measuring, harispe2015semantic}.

Recently, with the advances in AI, deep learning based approaches become increasingly popular~\cite{tai2015improved, he2016pairwise, wang2016sentence, devlin2018bert, tien2019sentence, reimers2019sentence, sun2020ernie}. 
Specifically, embeddings are broadly in use with a number of popular training schemes: continuous bag of words (CBOW) based training (e.g., doc2vec~\cite{le2014distributed, lau2016empirical}) and masked language models (MLM) based training~\cite{reimers2019sentence, devlin2018bert}

\noindent

\section{Design}
CTI reports have domain specific semantics. For example, `{\em IP}' and `{\em network}' have similar meanings, `{\em drop}' and `{\em payload}' have strong correlations. Such semantics can hardly be captured by general-purpose language models. This challenge can be overcome using domain specific corpus during training. 
Therefore, a straightforward proposal is to use masked language model to train on a large corpus of CTI reports and then use sentence embeddings in CTI search. Specifically, a malware IoC is described by some sentence(s). The search can simply look for CTI reports that contain similar sentence embeddings. However, such a proposal can hardly work because distinct behaviors by small nuance differences between {\bf B-1}, i.e., ``{\em use a dropper that has encrypted payload}'' and an entirely different behavior such as ``{\em drops an encrypted payload}'' will not be captured by embedding techniques. As of its ramifications, our later evaluation shows lower precision (i.e., false positives) by embedding techniques (Table~\ref{table:eval-precision-result}).

Furthermore, a CTI report may also describes a behavior using different sentence structures such that {\bf A-1} reads ``{\it The malware is an {\bf encrypted} and {\bf obfuscated} {\bf binary}, ..., it {\bf drops} a piece of shell code ...}'' which is more verbosely written. 
Although humans can easily determine that the corresponding malware has behavior {\bf B-1}, the syntactic-based analysis (e.g., syntactic dependency analysis) can also fail.  

We observe that the attention mechanism in Transformer models can capture (domain specific) semantic correlations between words.
For example, there are strong correlations between `\textit{dropper}' and ``\textit{encrypted payload}'' in \textbf{B-1} and two strong correlations in \textbf{B-3}: `\textit{collect}' - `\textit{host}' and `\textit{host}' - `\textit{IP address}'. 
These correlations are often not syntactic like verb-object relations.
We depict the detail on how attention mechanism works with those examples (\textbf{A1}-\textbf{A3} and \textbf{B1}-\textbf{B3}) in Appendix~\ref{section:more_on_motivation_examples}.

\begin{figure}[H]
    \centering
    {%
    \includegraphics[width=\linewidth]{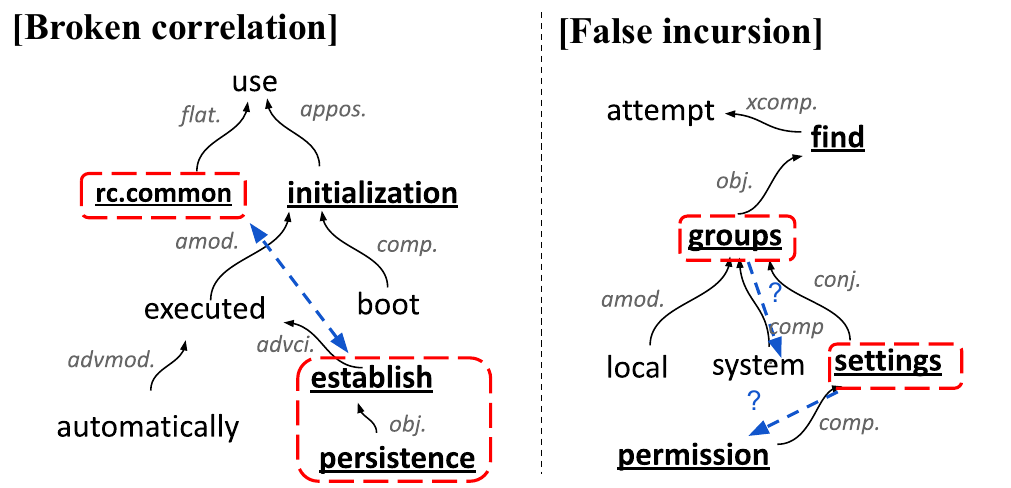}
    }
	\caption{Inaccuracies in use of dependency trees (from two sentences in Figure~\ref{fig:heat-map-example}). A dependency tree shows semantically correlated, but broken clauses (red boxes) due to no syntactic relation (left). Also it may incur false positive correlations (blue arrows) due to multiple equivalent neighbors (right).}
	\label{fig:example-dp-failing}
\end{figure}
\vspace{-2mm}

It is believed that self-attentions map the word-to-word correlations in domain specific semantics.
In this regard, our idea is to extract semantically structured graphs from text using self-attention maps.
To the best of our knowledge, it is novel methodology to use attention mechanism for a semantic dependency parsing.
Here, the graph construction is to traverse the sentence (i.e., set of words), prioritized by higher attention scores. 
Figure~\ref{fig:heat-map-example} illustrates how graphs can be constructed along with attention scores. 
Here, the key of semantic graph extraction is to abstract the core behaviors as a sub-graph form. 
All of above sentences are achieved abstractions by the self-attention guided exploration.

To compare attention maps with legacy (syntactic) dependency trees, we revisit two of above sentences with their parse trees (in Figure~\ref{fig:example-dp-failing}).
It shows that the parser fails to capture the long-distance correlation in a lengthy text (i.e., broken correlation).
The ramification is that it may need to include unnecessary words between two, i.e., incurring false positives.
Also, syntactic relations may cause to explore the obsolete paths (i.e., false incursion).

{\bf Training and Use of Attentions.}
The benefit of our method is that the model uses self-supervised training. 
As myriad of security problems, the lack of labeled dataset crucially harm the model performances. 
Albeit it is feasible to collect a large scale of CTI text corpus, we have no supervision for training (i.e., no pairwise ground truth).
In this regard, our method fits our domain problem in that it exploits the self-supervised learning. We hence use masked language model with a \textit{BERT}~\cite{devlin2018bert} architecture on the large-scale (8M words) CTI corpus (refer to Table~\ref{table:dataset-unlabeled}).

\vspace{-2mm}
\begin{figure}[H]
    \centering
    \vspace{3mm}
    {%
    \setlength{\fboxsep}{2pt}%
    \setlength{\fboxrule}{0.5pt}%
    \includegraphics[clip=true,trim=0mm 0mm 0mm 0mm,width=.9\linewidth]{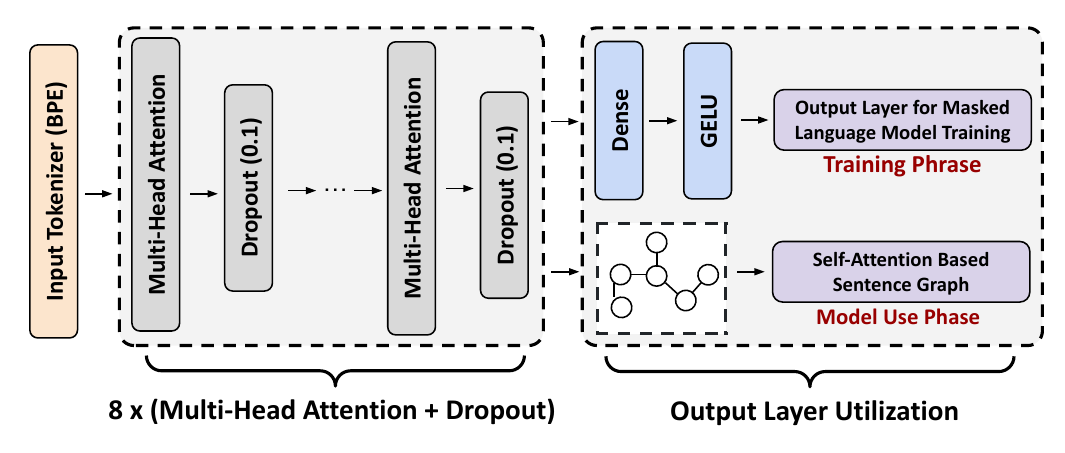}
    \vspace{-3mm}
    }%
     
	\caption{The utilization of self-attention for search.}
	\label{fig:overwiew-of-self-attention-layer}
\end{figure}
\vspace{-3mm}
 
The use of such model (i.e., pre-trained) is different from legacy that in LLMs (e.g., fine-tune) in that we construct the graphs by exploring the attention maps.
We leverage the self-attention scores to prioritize edges with higher attentions. 
In such method, it does not entail any supervised learning. 

\noindent

\begin{table*}[tb]
\centering
\caption{Effectiveness Evaluation (P stands for precision and R for recall)}
\label{table:eval-precision-result}
\begin{tabular}{|c|c|ccc|ccc|} 
\hline
\multicolumn{2}{|c|}{\multirow{2}{*}{\textbf{Type}}}    & \multicolumn{3}{c|}{\textbf{SP-EVAL-SET-1}}                                                     & \multicolumn{3}{c|}{\textbf{SP-EVAL-SET-1}}                                                      \\ 
\cline{3-8}
\multicolumn{2}{|c|}{}                                  & \multicolumn{1}{c|}{\textbf{P.}} & \multicolumn{1}{c|}{\textbf{R.}} & \textbf{F1}               & \multicolumn{1}{c|}{\textbf{P.}} & \multicolumn{1}{c|}{\textbf{R.}} & \textbf{F1}                \\ 
\hline\hline
\multicolumn{2}{|c|}{\textit{Simple Word Matching}}                     & 0.59                             & 0.97                             & 0.73                      & 0.60                             & 0.98                             & 0.75                       \\ 
\cline{1-2}
\multicolumn{2}{|c|}{\textit{Doc2Vec}~\cite{le2014distributed}}                           & 0.69                             & 0.77                             & 0.73                      & 0.66                             & 0.81                             & 0.73                       \\ 
\cline{1-2}
\multicolumn{2}{|c|}{\textit{Transformer}~\cite{reimers2019sentence}}                       & 0.61                             & 0.91                             & 0.73                      & 0.84                             & 0.68                             & 0.75                       \\ 
\cline{1-2}
\multicolumn{2}{|c|}{\textit{Transformer-Finetune}~\cite{turc2019well}}                  & 0.62                             & 0.89                             & 0.73                      & 0.65                             & 0.91                             & 0.76                       \\ 
\cline{1-2}
\multicolumn{2}{|c|}{\textit{Keyword Similarity 1}~\cite{mihalcea2006corpus}}                         & 0.73                             & 0.90                             & 0.80                      & 0.73                             & 0.88                             & 0.79                       \\ 
\cline{1-2}
\multicolumn{2}{|c|}{\textit{Keyword Similarity 2}~\cite{harispe2015semantic}}                         & 0.60                             & 0.95                             & 0.74                      & 0.55                             & 0.98                             & 0.71                       \\ 
\cline{1-2}
\multicolumn{2}{|c|}{\textit{Graph Isomorphism w/ Dependancy Parser}}                         & 0.75                             & 0.89                             & 0.81                      & 0.77                             & 0.88                             & 0.82   \\  
\cline{1-2}
\multicolumn{2}{|c|}{\textit{\textbf{Attention Graph Isomorphism} (Our)}}                         & 0.82                             & 0.93                             & \textbf{0.87}                      & 0.81                             & 0.89                             & \textbf{0.85}   \\     
\hline
\end{tabular}
\end{table*}

\smallskip
\noindent
{\bf Sub-graph Matching and Similarity Score.}
After graph constructions, we use the sub-graph matching algorithm and the similarity score computation. 
We consider two words $w_1$  and $w_2$ match if and only if $|e(w_1) - e(w_2)| < \tau$ where $e(w)$ denotes the embedding of a word $w$ and $\tau$ is a threshold.
With the definition of matching nodes, our challenge can hence be reduced to the {\em subgraph isomorphism problem}~\cite{sys1982subgraph}, which aims to find the largest isomorphic sub-graph of two un-directed graphs (Algorithm~\ref{alg:cap}).
The complexity of the problem is NP. However, since the graph for behavior description is small (10-15 nodes on average), the runtime is reasonable in practice, with our optimization of filtering irrelevant articles mentioned in Appendix~\ref{section:efficiency}. 

The similarity score of two isomorphic subgraphs $G_1$ and $G_2$ is hence computed as follows.

\vspace{-0.5cm}
$$\textrm{sim} (G_{1}, G_{2}) = \prod_{w_1, w_2 \in G_1 \times G_2} \Big[ \kappa (1-|e(w_1)-e(w_2)|)
\Big]$$ 

where $\kappa$ is a constant larger than 1, $e(w_1)$, and $e(w_2)$ the embeddings normalized to [0,1].
Note that $\kappa$ needs to be larger than 1 such that large isomorphic sub-graphs yield a larger similarity score. 

{\bf Implementation.}
We use 8 layers of multi-head attentions (of size 512) with a dropout on each layer (0.1 probability).
Preprocessing is largely standard with some domain specific normalization for IoC related artifacts, such as IP.
Specifically, we feed each CTI report under search to the model and acquire the self-attentions at the last layer. 
For each input text, we construct an {\em attention graph}, in which each node denotes a token. An edge is introduced between two nodes when their attention exceeds a threshold of 0.15. 
In sub-graph matching, we use 2.72, and 0.37 as $\kappa $ and $\tau$ threshold each. 
\section{Evaluation}

To train our models, we have collected 10,544 threat analysis articles from eight major security vendors (refer to Table~\ref{table:dataset-unlabeled}). 
The corpus contains 500K sentences and 8M words. 
For the input tokenizing for self-attention layers, we use {\em byte-pair encoding}~\cite{sennrich2015neural} and limit the number of tokens to 30,000 (originally, the dataset has 185K distinct words after lemmatization). 
We use \textit{BERT} for the self-attention layers (with 20\% of random masking and 2 epochs) and the \textit{Gensim} framework to train a \textit{word2vec} model with output vector size of 100 and 100 epochs for domain specific word embeddings.

\vspace{-0.1cm}
\begin{table}[H]
\centering
\caption{Pretraining Dataset}\label{table:dataset-unlabeled}
\vspace{-3mm}
\begin{tabular}{cccc} 
\hline
\textbf{Vendor}                                      & \begin{tabular}[c]{@{}c@{}}\textbf{\# of~}\\\textbf{Articles}\end{tabular} & \begin{tabular}[c]{@{}c@{}}\textbf{\# of }\\\textbf{Sents}\end{tabular} & \begin{tabular}[c]{@{}c@{}}\textbf{\# of }\\\textbf{Words}\end{tabular}  \\ 
\hline
\textit{FireEye}                                                                                                  & 843                                                                       & 50K                                                                      & 858K                                                                  \\
\rowcolor[rgb]{0.902,0.902,0.902} \textit{Fortinet}                                                            & 541                                                                       & 49K                                                                      & 645K                                                                  \\
\textit{IBM}                                                                                                & 926                                                                       & 43K                                                                      & 843K                                                                  \\
\rowcolor[rgb]{0.902,0.902,0.902} \textit{Kaspersky}                                                             & 1,441                                                                     & 50K                                                                      & 858K                                                                  \\
\textit{McAfee}                                                                                                & 626                                                                       & 38K                                                                      & 587K                                                                  \\
\rowcolor[rgb]{0.902,0.902,0.902} \textit{Palo Alto}                                                           & 641                                                                       & 58K                                                                      & 897K                                                                  \\
\textit{Symantec}                                                                                           & 177                                                                       & 16K                                                                      & 278K                                                                  \\
\rowcolor[rgb]{0.902,0.902,0.902} \textit{ESET}                                                             & 5,349                                                                     & 149K                                                                     & 3M                                                                \\ 
& \vspace{-3.0mm} \\ 
\hline
\textbf{Total}                                                                                                          & \textbf{10.5K}                                                           & \textbf{0.5M}                                                            & \textbf{8M}                                                       \\
\hline
\end{tabular}
\end{table}

Our evaluation answers follwing research questions: ({\bf R1}) what is the effectiveness of the proposed method compared to existing techniques; ({\bf R2}) how does the technique help real-world attack investigation; and ({\bf R3}) how efficient is the method.

\subsection{Effectiveness}
We devise a controlled experiment to compare the precision and recall of different methods.
We use the \textbf{SP-EVAL-SET-1}~\cite{mitreattack} and \textbf{SP-EVAL-SET-2}~\cite{capec} datasets.
Specifically, they provide attack behavior dictionaries such that for each threat behavior, they provide (i) a written description for the behavior and (ii) the associated real-world malware cases' descriptions.
For each behavior, we construct a dataset as follows. We include all the malware cases associated with the behavior (the true positives) and the same number of random cases from other behaviors.
In total, we test 423 behaviors from \textbf{SP-EVAL-SET-1}, 262 behaviors from \textbf{SP-EVAL-SET-2} and the aggregated number of cases are 14,096 and 2,002 for \textbf{SP-EVAL-SET-1} and \textbf{SP-EVAL-SET-2}, respectively.

We use the following baselines that are unsupervised learning based.

\begin{itemize}
  \setlength{\itemsep}{-1.2mm}
  \item \textit{\textbf{Word Matching}}: returning sentences based on the common words.
  \item \textit{\textbf{Doc2Vec}}: A sentence embedding technique based on the CBOW model~\cite{le2014distributed}. It returns sentences based on embedding similarities.
  \item \textit{\textbf{Transformer}}: Training a \textit{Transformer} model ~\cite{reimers2019sentence} from scratch and using sentence embedding similarity.
  \item \textit{\textbf{Transformer-Finetune}}:
  Fine-tuning a pre-trained   \textit{BERT} model~\cite{github2018bert} and using embedding similarity.
  \item \textit{\textbf{Keyword Similarity 1}}: A widely used text similarity based method~\cite{mihalcea2006corpus} using word weights.
  \item \textit{\textbf{Keyword Similarity 2}}: A recent work  prioritizing short texts, e.g., compact keywords~\cite{kenter2015short}.
  \item \textit{\textbf{Graph Isomorphism}}: Our methodology.
\end{itemize}

All these results require a threshold to determine the retrieved cases. We try many thresholds and report the best results. 
The results are presented in Table~\ref{table:eval-precision-result}.
Observe our method (the last row) achieves the best performance. It has highest F1 score than any other baselines.
Also observe that directly using sentence embeddings does not yield good results, neither do the keyword similarity based methods.

\noindent
{\bf Analysis of Failing Cases.}
Table~\ref{table:failing-cases-baseline} shows a few failing cases by the baselines.
In the first case, the embedding based methods yield the wrong results as the sentence embeddings are dominated by the verb such that the sentences with `{\it display}', `{\it find}' and `{\it identify}' are matched with the query sentence through the verb `{\it get}'.   
In comparison, the matching attention graphs by our method better disclose the essence.
In the second case, the embedding based methods focus too much on the verbs. 
The keyword based methods report the wrong results because they find three keyword matches. However, these keywords do not have the semantic correlations as those in the true positive.
We can observe the similar cases in the third.

\subsection{Use in Real-world Attack Forensics}

A critical task in forensics is to identify attack origins, that is, attributing attacks to their threat actors.
In this experiment, we randomly select a few recent attacks and assume a subset of behaviors are known beforehand. We then use them to search the corresponding CTI reports.
We use Google and IoC matching (similar to VirusTotal) as the baselines.

First, we collect additional CTI reports with explicit attack origin information and exclude all that overlap with the training set.
The collection contains 258 articles with 12 major actors. 
The details are in Table~\ref{table:dataset-labeled-2} (Appendix). 
We then randomly gather 10 real-world attacks in \textit{Mitre ATT\&CK}. We use their behavior descriptions to search the article pool. We consider the actor with the largest number of matching as the attack actor.

\begin{figure*}[tb]
    \centering
    \vspace{3mm}
    {\scriptsize \setstretch{1.1}
    \begin{tabular}[b]{cc}
     \textbf{Malware} & \textbf{Origin}\\ 
     \hline
     \textit{Winnti}        & \textit{Axiom}\\
     \textit{MessageTap}    & \textit{Axiom}\\
     \textit{DTrack}        & \textit{Lazarus}\\
     \textit{Hot C.}        & \textit{Lazarus}\\
     \textit{F. POS}        & \textit{FIN6}\\
     \textit{KerrDown}      & \textit{APT32}\\
     \textit{RDAT}          & \textit{Chry.}\\
     \textit{comRAT}        & \textit{Turla}\\
     \textit{RainyDay}      & \textit{Naikon}\\
     \textit{ServHelper}    & \textit{TA505}\\\hline
     \\
     \\
     \\
     \\
    
    \end{tabular}
    }
    \includegraphics[clip=true,trim=0mm 0mm 0mm 0mm,width=.255\linewidth]{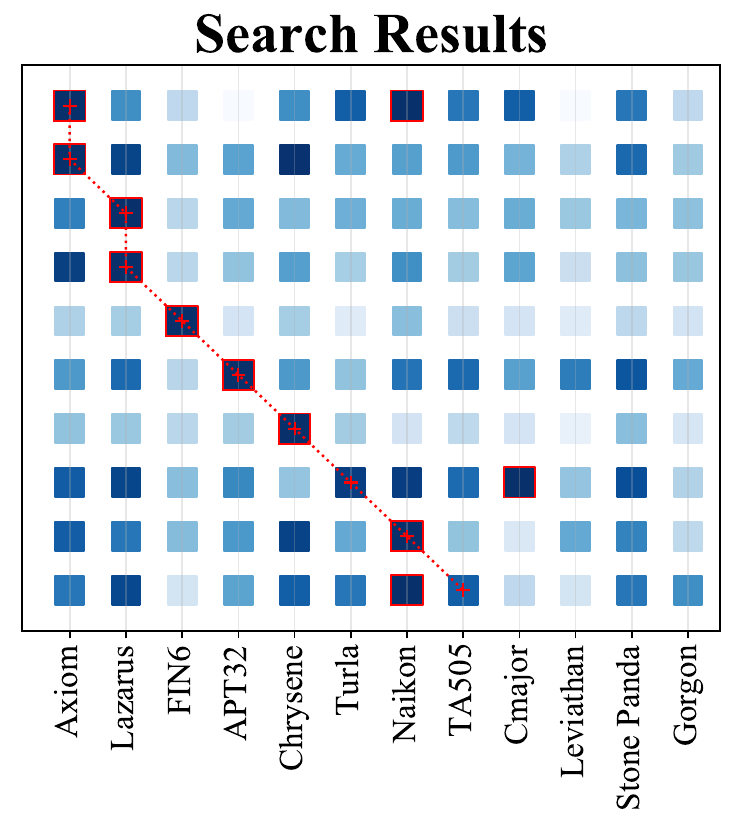}
    \includegraphics[clip=true,trim=0mm 0mm 0mm  0mm,width=.255\linewidth]{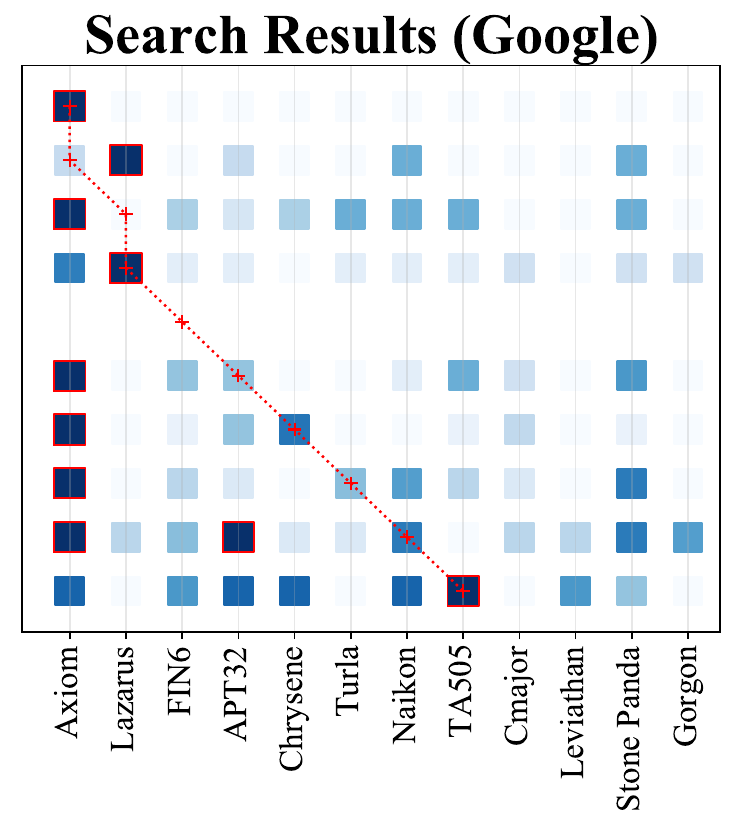}
    
    \vspace{-2mm}
       
	\caption{Search Result for Attack Origin Identification.
	In each figure, a red cross on the line {\footnotesize (\textcolor{red}{\textbf{+}---\textbf{+}---\textbf{+}})} denotes the target origin of a malware. In each row (i.e., a malware), the actor with the largest number of search results is marked by a red square {\footnotesize (\textcolor{red}{$\square$})}. Therefore, 
	a co-location of the two symbols tells the success of origin identification (\textcolor{red}{$\scalemath{0.6}{\mathrel{\young(+)}}$}).}
	\label{fig:search-result-origin-indentificaton}
\end{figure*}

Figure~\ref{fig:search-result-origin-indentificaton} represents the results.
Our method successfully identifies 8 correct origins out of 10 cases, whereas Google identifies 3 correct answers.

To compare with IoC matching, 
we manually extract all IoC artifacts from the threat reports. It is common practice that authors attach such information), including the following types: \textit{URL}, \textit{IP}, \textit{Hash}, \textit{CVE}, \textit{Registry}, \textit{File}~\cite{github2018openioc}. We exclude trivial \textit{Windows} executable file names (e.g., \textrm{cmd.exe}) which cause a high volume of false positives.
We then use exact matches of IoCs in the experiment.
As a result, only two attacks ({\it Winnti} and {\it RDAT}) can be attributed to their origins.

We also test the SOTA internet-connected LLM, namely GPT-4 (i.e., Bing Chat) by prompting to extract related articles. 
GPT-4 only answers for 5 origins. Among those 5, it can correctly attribute 3; \textit{Dtrack}, \textit{HotCroissant} and \textit{KerrDown}.  

\begin{table*}[tb]
\caption{ Failing cases of baselines.
Matching background colors denote matching words (based on embeddings). 
The underlined words in the query and the sentence returned by our method form the matching attention graphs. Our method returns the correct CTIs in all these cases. 
}
\vspace{-2mm}
\label{table:failing-cases-baseline}

\begin{tabular}{|l|} 
\hline
\textbf{[Query]} \textit{\colorbox{orange!20}{\underline{\textbf{get}}} \colorbox{blue!20}{information} about \colorbox{yellow!20}{\underline{\textbf{running}}} \underline{\textbf{processes}} on \colorbox{red!20}{system}}                                                                                                                                                                                                                                             \\ 
\hline
\textbf{\textbf{[D2V]}}  \textit{Sykipot may use netstat to \colorbox{orange!20}{display} active network connections}                  
\\
\vspace{-2mm}
\\
\textbf{[Transformer]}  \textit{GravityRAT uses \textit{netstat} to \colorbox{orange!20}{find} open ports on victim’s \colorbox{red!20}{machine}}                                                                                                                                                                                                                   \\
\textbf{\textbf{[Transformer]}}  \textit{Get2 has ability to \colorbox{orange!20}{identify} current username of infected host}                                                                                                      
\\
\\

\textbf{\textbf{[Our Search Result]~}}  \textit{PowerShower has ability to ... module to \underline{\textbf{retrieve}} list of \underline{\textbf{active}} \underline{\textbf{processes}}}        
\\
\vspace{-2mm}
\\

\hline
\hline

\textbf{\textbf{[Query]}} \textit{\colorbox{orange!20}{\underline{\textbf{execute}}} their own malicious \colorbox{blue!20}{\underline{\textbf{payloads}}} by \underline{\textbf{hijacking}} \colorbox{yellow!20}{\underline{\textbf{library}}} manifest used to load DLLs}                                                                                                                                                                                         \\ 
\hline
\textbf{\textbf{[D2V, Transformer]}}  \textit{APT28 has used tools to \colorbox{orange!20}{perform} keylogging}                                                                                                                                                                                                                                                        \\
\textbf{\textbf{\textbf{\textbf{[D2V]}}}}  \textit{BADNEWS is capable of \colorbox{orange!20}{executing} commands via cmd.exe}       \\
\vspace{-2mm}
\\
\textbf{\textbf{[Keyword Similarity]~}}  \textit{SeaDuke uses \colorbox{yellow!20}{module} to \colorbox{orange!20}{execute} Mimikatz with \colorbox{blue!20}{PowerShell} to}  \\
\textit{perform Pass Ticket} \\
\textbf{\textbf{[Keyword Similarity]~}}  \textit{PowerSploit \colorbox{yellow!20}{modules} are written in and \colorbox{orange!20}{executed} via \colorbox{blue!20}{PowerShell}}                                              
\\
\\
\textbf{\textbf{[Our Search Result]~}}  \textit{HyperBro has used legitimate application to \textbf{\underline{sideload} \underline{DLL}} to decrypt} \\
\textit{decompress and \textbf{\underline{run} \underline{payload}}.} 

\\
\hline
\hline
\begin{tabular}[c]{@{}l@{}}\textbf{\textbf{[Query]}} \textit{with no prior \colorbox{blue!20}{knowledge} of legitimate \colorbox{yellow!20}{credentials} within \colorbox{red!20}{system} or environment,} \\  \textit{\colorbox{orange!20}{\underline{\textbf{guess}}} \underline{\textbf{passwords}} to attempt \underline{\textbf{access}} to \underline{\textbf{accounts}}}\end{tabular}                                                                                                                                                     \\ 
\hline
\begin{tabular}[c]{@{}l@{}}\textbf{\textbf{[D2V, Transformer ]}}~\textit{Zeus Panda \colorbox{orange!20}{checks} to see if anti virus anti spyware or firewall products} \\ \textit{are installed in victim’s environment}  \end{tabular} 
\\
\vspace{-2mm}
\\
\begin{tabular}[c]{@{}l@{}}\textbf{\textbf{[Keyword Similarity]~}} \textit{Agent Tesla can collect \colorbox{red!20}{system}'s computer name and also has capability to} \\ \textit{collect \colorbox{blue!20}{information} on processor memory and video \colorbox{yellow!20}{card}} \end{tabular}                 
\\
\\
\textbf{\textbf{[Our Search Result]~}} \textit{SpeakUp can perform \underline{\textbf{bruteforce}} using predefined list of usernames }\\ \textit{and \underline{\textbf{passwords}} in attempt to \underline{\textbf{log-in}} to \underline{\textbf{administrative}} panels} 
\\

\hline

\end{tabular}
\end{table*}

\subsection{Efficiency}
We have implemented non-lossy search optimizations; (i) graph caching and (ii) sentences clustering and evaluated the runtime efficiency of our method. Our optimizations, evaluation and system specification details can be found in Appendix~\ref{section:efficiency}.

\begin{table}[H]
\centering
\caption{Result on Efficiency Test}
\label{table:efficiency-test}
\begin{tabular}{|c|c|c|c|} 
\hline
\multirow{2}{*}{\textbf{Type}}                                         & \multicolumn{3}{c|}{\textbf{Search Space}}   \\ 
\cline{2-4}
                                                                       & \textbf{20K} & \textbf{50K} & \textbf{100K}  \\ 
\hhline{|====|}
\begin{tabular}[c]{@{}c@{}}\textbf{Baseline}\\(matching)\end{tabular}  & 20s          & 53s          & 1m45s          \\ 
\hline
\begin{tabular}[c]{@{}c@{}}\textbf{Our System}\\(optimized)\end{tabular} & 09s          & 28s          & 57s            \\
\hline
\end{tabular}
\end{table}

We compare our system (after fully optimized) to the simplest word matching in efficiency. The test performs the search query of 10 words by varying search space sizes.
The baseline includes a text preprocessing and word-to-word comparison within a pair of two sentences. 

Before any optimizations, a raw implementation of the graph isomorphism costs $\sim$17m with 20K search space, $\sim$4h with 100K (Table~\ref{table:efficiency-test-full}).
However, our non-lossy optimization drastically reduces search times as shown in Table~\ref{table:efficiency-test} that become comparable to the baseline. 

\section{Discussion}
\textbf{Word Embedding.}
One can benefit from contextualized word embedding. 
Also, the \textit{Transformer} model contains pre-trained token embedding that can be used to measure word to word distance. It is worth considering a use of such embedding methodologies. 
In our study, we use \textit{word2vec} in that attack describing terms (e.g., exploit, encrypt) tend to be absolute in different contexts.

\textbf{Traversal with Self-Attentions.}
While we use a threshold to construct graphs from text (Algorithm~\ref{alg:cap}), the methodology is not limited in general. For example, one can use a traversal algorithm prioritized by attention scores.
Attention map may also help to build syntax parsers as it retrieve semantic correlations in sentences.

\textbf{Dataset Release.}
We publish our dataest\footnote{\href{https://github.com/cwbae10-purdue/CTI-EACL24.git}{https://github.com/cwbae10-purdue/cti-eacl24.git}}. 
Any following work must be only on research purposes.
It is worth noting that our dataset is a collection of open articles from various vendors. 
As our dataset is up to year of 2023, one may need to reproduce the collection by following the instruction.

\section{Conclusion}
We propose a novel CTI search method using attention graph isomorphism. 
We have shown that our method improves the effectiveness of CTI search for comparative evaluations. Our case study also shows that it drastically improves attack origin identification.
Our technique can correctly attribute 8 of 10 recent attacks while Google only attributes 3.

\section{Limitations} 
Since our system resorts to  word-level embeddings, it has difficulty handling cases in which a word is equivalent to a phrase.
For example, \textit{“obfuscate”} and \textit{“make it difficult to understand”} are semantically similar. But such similarity may not be captured by our technique.
We speculate with a large corpus, the embeddings by transformer can better consider the context and hence capture the similarity.

\newpage

\section{Ethnics Statements} 

Our system resorts to cyber threat intelligence (CTI) dataset. 
This may impose the risk factors as follows;

\begin{itemize}
  \setlength{\itemsep}{-1.2mm}
  \item \textit{\textbf{Possible Exposure of Threat Knowledge}}: As dataset comprise threat analysis articles, it may contain potential risks to be abused.
  \item \textit{\textbf{Adversarial Use of Knowledge}}: Attackers may use the information retrieval system to operate advanced attacks or avoid possible defenses assisted by threat intelligence.
  \item \textit{\textbf{Concerns on Privacy Information}}: Threat knowledge contained in the dataset may hold real cyberattack cases. It may contain state-wide or private damages or loss which can lead to violation of privacy. 
\end{itemize}

We have only used the dataest for research purposes (i.e., textual analysis). All reproduced work from our dataset must be on same objectives. 
In this regard, good practices as below need to be observed; 

\begin{itemize}
  \setlength{\itemsep}{-1.2mm}
  \item \textit{\textbf{Avoiding Use of Recent Critical Knowledge}}: One must refrain from the use of on-going (or recent) threat information as it might aggravate the circumstances.
  \item \textit{\textbf{Avoiding Use of Effective Vulnerabilities}}: If threat knowledge is still effective (e.g., before patch), one must not include such information.
  \item \textit{\textbf{Excluding Privacy Information on Damages}}: One must refrain from including privacy damages (e.g., loss/damage of specific institutions) to technical articles (e.g., case studies). It may contain privacy information.
\end{itemize}

\nocite{*}

\bibliography{anthology}

\begin{thebibliography}{59}
\expandafter\ifx\csname natexlab\endcsname\relax\def\natexlab#1{#1}\fi

\bibitem[{BERT()}]{github2018bert}
BERT.
\newblock \url{https://github.com/google-research/bert}.
\newblock Accessed: 2022-06-20.

\bibitem[{Budanitsky and Hirst(2006)}]{budanitsky2006evaluating}
Alexander Budanitsky and Graeme Hirst. 2006.
\newblock Evaluating wordnet-based measures of lexical semantic relatedness.
\newblock \emph{Computational linguistics}, 32(1):13--47.

\bibitem[{CAPEC()}]{capec}
CAPEC.
\newblock \url{https://capec.mitre.org/}.
\newblock Accessed: 2022-06-20.

\bibitem[{Catakoglu et~al.(2016)Catakoglu, Balduzzi, and
  Balzarotti}]{catakoglu2016automatic}
Onur Catakoglu, Marco Balduzzi, and Davide Balzarotti. 2016.
\newblock Automatic extraction of indicators of compromise for web
  applications.
\newblock In \emph{Proceedings of the 25th international conference on world
  wide web}, pages 333--343.

\bibitem[{Chen and Manning(2014)}]{chen2014fast}
Danqi Chen and Christopher~D Manning. 2014.
\newblock A fast and accurate dependency parser using neural networks.
\newblock In \emph{Proceedings of the 2014 conference on empirical methods in
  natural language processing (EMNLP)}, pages 740--750.

\bibitem[{Choshen et~al.(2019)Choshen, Eldad, Hershcovich, Sulem, and
  Abend}]{choshen2019language}
Leshem Choshen, Dan Eldad, Daniel Hershcovich, Elior Sulem, and Omri Abend.
  2019.
\newblock The language of legal and illegal activity on the darknet.
\newblock \emph{arXiv preprint arXiv:1905.05543}.

\bibitem[{Corley and Mihalcea(2005)}]{corley2005measuring}
Courtney~D Corley and Rada Mihalcea. 2005.
\newblock Measuring the semantic similarity of texts.
\newblock In \emph{Proceedings of the ACL workshop on empirical modeling of
  semantic equivalence and entailment}, pages 13--18.

\bibitem[{Croce et~al.(2011)Croce, Moschitti, and Basili}]{croce2011structured}
Danilo Croce, Alessandro Moschitti, and Roberto Basili. 2011.
\newblock Structured lexical similarity via convolution kernels on dependency
  trees.
\newblock In \emph{Proceedings of the 2011 Conference on Empirical Methods in
  Natural Language Processing}, pages 1034--1046.

\bibitem[{Devlin et~al.(2018)Devlin, Chang, Lee, and
  Toutanova}]{devlin2018bert}
Jacob Devlin, Ming-Wei Chang, Kenton Lee, and Kristina Toutanova. 2018.
\newblock Bert: Pre-training of deep bidirectional transformers for language
  understanding.
\newblock \emph{arXiv preprint arXiv:1810.04805}.

\bibitem[{{Economictimes}(2019)}]{economictimes2019dtrack}
{Economictimes}. 2019.
\newblock \href
  {https://economictimes.indiatimes.com/magazines/panache/what-is-dtrack-the-spytool-that-is-to-blame-for-attacks-on-indian-financial-institutions/articleshow/71706234.cms}
  {What is dtrack: North korean virus being used to hack atms to nuclear power
  plant in india}.
\newblock Published: 2019-10-22.

\bibitem[{Ferguson and Kaminsky(2008)}]{ferguson2008reverse}
Justin Ferguson and Dan Kaminsky. 2008.
\newblock \emph{Reverse engineering code with IDA Pro}.
\newblock Syngress.

\bibitem[{Gao et~al.(2021)Gao, Shao, Liu, Xiao, Qin, Xu, Mittal, Kulkarni, and
  Song}]{gao2021enabling}
Peng Gao, Fei Shao, Xiaoyuan Liu, Xusheng Xiao, Zheng Qin, Fengyuan Xu, Prateek
  Mittal, Sanjeev~R Kulkarni, and Dawn Song. 2021.
\newblock Enabling efficient cyber threat hunting with cyber threat
  intelligence.
\newblock In \emph{2021 IEEE 37th International Conference on Data Engineering
  (ICDE)}, pages 193--204. IEEE.

\bibitem[{Harispe et~al.(2015)Harispe, Ranwez, Janaqi, and
  Montmain}]{harispe2015semantic}
S{\'e}bastien Harispe, Sylvie Ranwez, Stefan Janaqi, and Jacky Montmain. 2015.
\newblock Semantic similarity from natural language and ontology analysis.
\newblock \emph{Synthesis Lectures on Human Language Technologies},
  8(1):1--254.

\bibitem[{He and Lin(2016)}]{he2016pairwise}
Hua He and Jimmy Lin. 2016.
\newblock Pairwise word interaction modeling with deep neural networks for
  semantic similarity measurement.
\newblock In \emph{Proceedings of the 2016 conference of the north American
  chapter of the Association for Computational Linguistics: human language
  technologies}, pages 937--948.

\bibitem[{Husari et~al.(2017)Husari, Al-Shaer, Ahmed, Chu, and
  Niu}]{husari2017ttpdrill}
Ghaith Husari, Ehab Al-Shaer, Mohiuddin Ahmed, Bill Chu, and Xi~Niu. 2017.
\newblock Ttpdrill: Automatic and accurate extraction of threat actions from
  unstructured text of cti sources.
\newblock In \emph{Proceedings of the 33rd Annual Computer Security
  Applications Conference}, pages 103--115.

\bibitem[{IDA()}]{hexrays}
IDA.
\newblock \url{https://hex-rays.com/}.
\newblock Accessed: 2022-06-20.

\bibitem[{{INDIA TODAY}(2019)}]{indiatoday2019dtrack}
{INDIA TODAY}. 2019.
\newblock \href
  {https://www.indiatoday.in/india/story/kudankulam-nuclear-power-plant-dtrack-north-korea-atms-1614200-2019-10-30/}
  {What is dtrack: North korean virus being used to hack atms to nuclear power
  plant in india}.
\newblock Published: 2019-10-30.

\bibitem[{{IOC Parser}()}]{github2018openioc}
{IOC Parser}.
\newblock \url{https://github.com/PaloAltoNetworks/ioc-parser}.
\newblock Accessed: 2022-06-20.

\bibitem[{Islam and Inkpen(2008)}]{islam2008semantic}
Aminul Islam and Diana Inkpen. 2008.
\newblock Semantic text similarity using corpus-based word similarity and
  string similarity.
\newblock \emph{ACM Transactions on Knowledge Discovery from Data (TKDD)},
  2(2):1--25.

\bibitem[{Jin et~al.(2022)Jin, Jang, Lee, Shin, and Chung}]{jin2022shedding}
Youngjin Jin, Eugene Jang, Yongjae Lee, Seungwon Shin, and Jin-Woo Chung. 2022.
\newblock Shedding new light on the language of the dark web.
\newblock \emph{arXiv preprint arXiv:2204.06885}.

\bibitem[{Kenter and De~Rijke(2015)}]{kenter2015short}
Tom Kenter and Maarten De~Rijke. 2015.
\newblock Short text similarity with word embeddings.
\newblock In \emph{Proceedings of the 24th ACM international on conference on
  information and knowledge management}, pages 1411--1420.

\bibitem[{Khandpur et~al.(2017)Khandpur, Ji, Jan, Wang, Lu, and
  Ramakrishnan}]{khandpur2017crowdsourcing}
Rupinder~Paul Khandpur, Taoran Ji, Steve Jan, Gang Wang, Chang-Tien Lu, and
  Naren Ramakrishnan. 2017.
\newblock Crowdsourcing cybersecurity: Cyber attack detection using social
  media.
\newblock In \emph{Proceedings of the 2017 ACM on Conference on Information and
  Knowledge Management}, pages 1049--1057.

\bibitem[{Lau and Baldwin(2016)}]{lau2016empirical}
Jey~Han Lau and Timothy Baldwin. 2016.
\newblock An empirical evaluation of doc2vec with practical insights into
  document embedding generation.
\newblock \emph{arXiv preprint arXiv:1607.05368}.

\bibitem[{Le and Mikolov(2014)}]{le2014distributed}
Quoc Le and Tomas Mikolov. 2014.
\newblock Distributed representations of sentences and documents.
\newblock In \emph{International conference on machine learning}, pages
  1188--1196. PMLR.

\bibitem[{Liao et~al.(2016)Liao, Yuan, Wang, Li, Xing, and
  Beyah}]{liao2016acing}
Xiaojing Liao, Kan Yuan, XiaoFeng Wang, Zhou Li, Luyi Xing, and Raheem Beyah.
  2016.
\newblock Acing the ioc game: Toward automatic discovery and analysis of
  open-source cyber threat intelligence.
\newblock In \emph{Proceedings of the 2016 ACM SIGSAC Conference on Computer
  and Communications Security}, pages 755--766.

\bibitem[{Lopez-Gazpio et~al.(2019)Lopez-Gazpio, Maritxalar, Lapata, and
  Agirre}]{lopez2019word}
Inigo Lopez-Gazpio, Montse Maritxalar, Mirella Lapata, and Eneko Agirre. 2019.
\newblock Word n-gram attention models for sentence similarity and inference.
\newblock \emph{Expert Systems with Applications}, 132:1--11.

\bibitem[{Malpedia()}]{malpedia}
Malpedia.
\newblock \url{https://malpedia.caad.fkie.fraunhofer.de/}.
\newblock Accessed: 2022-06-20.

\bibitem[{Melamud et~al.(2016)Melamud, Goldberger, and
  Dagan}]{melamud2016context2vec}
Oren Melamud, Jacob Goldberger, and Ido Dagan. 2016.
\newblock context2vec: Learning generic context embedding with bidirectional
  lstm.
\newblock In \emph{Proceedings of the 20th SIGNLL conference on computational
  natural language learning}, pages 51--61.

\bibitem[{Mihalcea et~al.(2006)Mihalcea, Corley, Strapparava
  et~al.}]{mihalcea2006corpus}
Rada Mihalcea, Courtney Corley, Carlo Strapparava, et~al. 2006.
\newblock Corpus-based and knowledge-based measures of text semantic
  similarity.
\newblock In \emph{Aaai}, volume~6, pages 775--780.

\bibitem[{Milajerdi et~al.(2019)Milajerdi, Gjomemo, Eshete, Sekar, and
  Venkatakrishnan}]{milajerdi2019holmes}
Sadegh~M Milajerdi, Rigel Gjomemo, Birhanu Eshete, Ramachandran Sekar, and
  VN~Venkatakrishnan. 2019.
\newblock Holmes: real-time apt detection through correlation of suspicious
  information flows.
\newblock In \emph{2019 IEEE Symposium on Security and Privacy (SP)}, pages
  1137--1152. IEEE.

\bibitem[{Miller(1995)}]{miller1995wordnet}
George~A Miller. 1995.
\newblock Wordnet: a lexical database for english.
\newblock \emph{Communications of the ACM}, 38(11):39--41.

\bibitem[{{Mitre ATTACK}()}]{mitreattack}
{Mitre ATTACK}.
\newblock \url{https://attack.mitre.org/}.
\newblock Accessed: 2022-06-20.

\bibitem[{NLTK()}]{nltk}
NLTK.
\newblock \url{https://www.nltk.org/}.
\newblock Accessed: 2022-06-20.

\bibitem[{Oktavianto and Muhardianto(2013)}]{oktavianto2013cuckoo}
Digit Oktavianto and Iqbal Muhardianto. 2013.
\newblock \emph{Cuckoo malware analysis}.
\newblock Packt Publishing Ltd.

\bibitem[{Rahutomo et~al.(2012)Rahutomo, Kitasuka, and
  Aritsugi}]{rahutomo2012semantic}
Faisal Rahutomo, Teruaki Kitasuka, and Masayoshi Aritsugi. 2012.
\newblock Semantic cosine similarity.
\newblock In \emph{The 7th international student conference on advanced science
  and technology ICAST}, volume~4, page~1.

\bibitem[{Ramage et~al.(2009)Ramage, Rafferty, and Manning}]{ramage2009random}
Daniel Ramage, Anna~N Rafferty, and Christopher~D Manning. 2009.
\newblock Random walks for text semantic similarity.
\newblock In \emph{Proceedings of the 2009 workshop on graph-based methods for
  natural language processing (TextGraphs-4)}, pages 23--31.

\bibitem[{Ramos et~al.(2003)}]{ramos2003using}
Juan Ramos et~al. 2003.
\newblock Using tf-idf to determine word relevance in document queries.
\newblock In \emph{Proceedings of the first instructional conference on machine
  learning}, volume 242, pages 29--48. New Jersey, USA.

\bibitem[{Rao et~al.(2019)Rao, Liu, Tay, Yang, Shi, and Lin}]{rao2019bridging}
Jinfeng Rao, Linqing Liu, Yi~Tay, Wei Yang, Peng Shi, and Jimmy Lin. 2019.
\newblock Bridging the gap between relevance matching and semantic matching for
  short text similarity modeling.
\newblock In \emph{Proceedings of the 2019 Conference on Empirical Methods in
  Natural Language Processing and the 9th International Joint Conference on
  Natural Language Processing (EMNLP-IJCNLP)}, pages 5370--5381.

\bibitem[{Reimers and Gurevych(2019)}]{reimers2019sentence}
Nils Reimers and Iryna Gurevych. 2019.
\newblock Sentence-bert: Sentence embeddings using siamese bert-networks.
\newblock \emph{arXiv preprint arXiv:1908.10084}.

\bibitem[{Securelist(2018)}]{securelist2018applejeus}
Securelist. 2018.
\newblock \href {https://securelist.com/operation-applejeus/87553/} {Operation
  applejeus: Lazarus hits cryptocurrency exchange with fake installer and macos
  malware}.
\newblock Published: 2018-08-23.

\bibitem[{Sennrich et~al.(2015)Sennrich, Haddow, and
  Birch}]{sennrich2015neural}
Rico Sennrich, Barry Haddow, and Alexandra Birch. 2015.
\newblock Neural machine translation of rare words with subword units.
\newblock \emph{arXiv preprint arXiv:1508.07909}.

\bibitem[{Song et~al.(2019)Song, Tan, Qin, Lu, and Liu}]{song2019mass}
Kaitao Song, Xu~Tan, Tao Qin, Jianfeng Lu, and Tie-Yan Liu. 2019.
\newblock Mass: Masked sequence to sequence pre-training for language
  generation.
\newblock \emph{arXiv preprint arXiv:1905.02450}.

\bibitem[{Sun et~al.(2020)Sun, Wang, Li, Feng, Tian, Wu, and
  Wang}]{sun2020ernie}
Yu~Sun, Shuohuan Wang, Yukun Li, Shikun Feng, Hao Tian, Hua Wu, and Haifeng
  Wang. 2020.
\newblock Ernie 2.0: A continual pre-training framework for language
  understanding.
\newblock In \emph{Proceedings of the AAAI Conference on Artificial
  Intelligence}, volume~34, pages 8968--8975.

\bibitem[{Sys et~al.(1982)}]{sys1982subgraph}
Maciej~M Sys et~al. 1982.
\newblock The subgraph isomorphism problem for outerplanar graphs.
\newblock \emph{Theoretical Computer Science}, 17(1):91--97.

\bibitem[{Tai et~al.(2015)Tai, Socher, and Manning}]{tai2015improved}
Kai~Sheng Tai, Richard Socher, and Christopher~D Manning. 2015.
\newblock Improved semantic representations from tree-structured long
  short-term memory networks.
\newblock \emph{arXiv preprint arXiv:1503.00075}.

\bibitem[{TechNadu(2019)}]{tech2019indian}
TechNadu. 2019.
\newblock \href
  {https://www.technadu.com/lazarus-group-new-banking-malware-against-indian-banks/80747/}
  {The lazarus group is using a new banking malware against indian banks}.

\bibitem[{Tien et~al.(2019)Tien, Le, Tomohiro, and Tatsuya}]{tien2019sentence}
Nguyen~Huy Tien, Nguyen~Minh Le, Yamasaki Tomohiro, and Izuha Tatsuya. 2019.
\newblock Sentence modeling via multiple word embeddings and multi-level
  comparison for semantic textual similarity.
\newblock \emph{Information Processing \& Management}, 56(6):102090.

\bibitem[{TrendMicro(2018)}]{trend2018ratankba}
TrendMicro. 2018.
\newblock \href
  {https://www.trendmicro.com/en_us/research/18/a/lazarus-campaign-targeting-cryptocurrencies-reveals-remote-controller-tool-evolved-ratankba.html/}
  {Lazarus campaign uses remote tools, ratankba, and more}.
\newblock Published: 2018-01-24.

\bibitem[{Turc et~al.(2019)Turc, Chang, Lee, and Toutanova}]{turc2019well}
Iulia Turc, Ming-Wei Chang, Kenton Lee, and Kristina Toutanova. 2019.
\newblock Well-read students learn better: On the importance of pre-training
  compact models.
\newblock \emph{arXiv preprint arXiv:1908.08962}.

\bibitem[{US-CERT(2017)}]{uscert2017hiddencobra}
US-CERT. 2017.
\newblock \href {https://www.cisa.gov/uscert/ncas/alerts/TA17-318A/} {Hidden
  cobra – north korean remote administration tool: Fallchill}.
\newblock Published: 2018-08-23.

\bibitem[{Vaswani et~al.(2017)Vaswani, Shazeer, Parmar, Uszkoreit, Jones,
  Gomez, Kaiser, and Polosukhin}]{vaswani2017attention}
Ashish Vaswani, Noam Shazeer, Niki Parmar, Jakob Uszkoreit, Llion Jones,
  Aidan~N Gomez, {\L}ukasz Kaiser, and Illia Polosukhin. 2017.
\newblock Attention is all you need.
\newblock \emph{Advances in neural information processing systems}, 30.

\bibitem[{VirusTotal()}]{virustotal}
VirusTotal.
\newblock \url{https://www.virustotal.com/}.
\newblock Accessed: 2022-06-20.

\bibitem[{Wang et~al.(2020)Wang, Liao, Qin, and Wang}]{wang2020into}
Peng~Wang Wang, Xiaojing~Liao Liao, Yue Qin, and XiaoFeng Wang. 2020.
\newblock Into the deep web: Understanding e-commercefraud from autonomous chat
  with cybercriminals.
\newblock In \emph{Proceedings of the ISOC Network and Distributed System
  Security Symposium (NDSS), 2020}.

\bibitem[{Wang et~al.(2016)Wang, Mi, and Ittycheriah}]{wang2016sentence}
Zhiguo Wang, Haitao Mi, and Abraham Ittycheriah. 2016.
\newblock Sentence similarity learning by lexical decomposition and
  composition.
\newblock \emph{arXiv preprint arXiv:1602.07019}.

\bibitem[{Yenicelik et~al.(2020)Yenicelik, Schmidt, and
  Kilcher}]{yenicelik2020does}
David Yenicelik, Florian Schmidt, and Yannic Kilcher. 2020.
\newblock How does bert capture semantics? a closer look at polysemous words.
\newblock In \emph{Proceedings of the Third BlackboxNLP Workshop on Analyzing
  and Interpreting Neural Networks for NLP}, pages 156--162.

\bibitem[{You and Yim(2010)}]{you2010malware}
Ilsun You and Kangbin Yim. 2010.
\newblock Malware obfuscation techniques: A brief survey.
\newblock In \emph{2010 International conference on broadband, wireless
  computing, communication and applications}, pages 297--300. IEEE.

\bibitem[{ZDNet(2019)}]{zdnet2019atm}
ZDNet. 2019.
\newblock \href
  {https://www.zdnet.com/article/new-north-korean-malware-targeting-atms-spotted-in-india/}
  {New north korean malware targeting atms spotted in india}.

\bibitem[{Zhanna Malekos~Smith(2020)}]{zhanna2020hiddencost}
Eugenia Lostri James A.~Lewis Zhanna Malekos~Smith. 2020.
\newblock The hidden costs of cybercrime.
\newblock Accessed: 2017-11-14.

\bibitem[{Zhu and Dumitras(2018)}]{zhu2018chainsmith}
Ziyun Zhu and Tudor Dumitras. 2018.
\newblock Chainsmith: Automatically learning the semantics of malicious
  campaigns by mining threat intelligence reports.
\newblock In \emph{2018 IEEE European Symposium on Security and Privacy
  (EuroS\&P)}, pages 458--472. IEEE.

\end{thebibliography}
\bibliographystyle{acl_natbib}

\appendix
\newpage
\section{Appendix}

\subsection{Ablation Study}

We conduct an ablation study in that; 

(i) we first eliminate a self-attention based graph builder and connect all possible pairs (make sentence a fully-connected-graph) and run isomorphic sub-graph discovery to get similarity score, 

(ii) we do not use \textit{Word2Vec} model so that isomorphic sub-graph cannot tolerate similar, but different words, therefore, it is only able to match same words.

\begin{table}[H]
\centering
\caption{Result on Ablation Study}
\label{table:ablation-test}
\begin{tabular}{|c|ccc|c|} 
\hline
\textbf{Type}                    & \multicolumn{1}{c|}{\textbf{P.}} & \multicolumn{1}{c|}{\textbf{R.}} & \textbf{F1} & \textbf{$\Delta$ F1}  \\ 
\hline \hline
w/o Self-Attention & .75                              & .78                              & .76         & -12.64\%              \\ 
\cline{1-1}
w/o Word2Vec                     & .82                              & .82                              & .82         & -5.75\%              \\
\hline
\end{tabular}
\end{table}

Under this configuration, we revisit the search performance evaluation on \textbf{SP-EVAL-SET-1}. 
For each modification, we set the threshold value again which maximizes the \textbf{F1}-score.  
Table~\ref{table:ablation-test} shows the result of ablation test. 
The "\textbf{Loss of F1}" column says the loss of \textbf{F1} score in percentile from our original result (note that our original method scores 87\% of \textbf{F1}-score). 

\subsection{Search Query Time}
\label{section:efficiency}
Before we measure the query-time performance, we introduce two optimization techniques; (i) graph caching (GC) and (ii) sentences clustering (SC);

We notice that our self-attention based graph builder is a bottleneck against time performance, in average, costs 0.5s for a single sentence (while isomorphic sub-graph discovery module takes 5ms in average which is almost zero relatively). 
However, graph building for search space sentences does not need to be processed on-the-fly. Therefore, we use the graph caching (GC) optimization where we pre-build graphs from the sentences and cache to the database. 

Also, we use a lossless optimization method, namely, sentence clustering (SC) in that we filter our any of non-related sentences have no synonym from the query sentence (recall that synonym is defined as two words those embedding vectors are within the constant $\tau$-distance). 
For that, we pre-build a map from every word in dictionary (185K words) to sentences in search space (i.e., sentence clustering by word that includes all sentences which hold at least one of its synonyms). When a query sentence is given, we extract words from it and load their sentence-clusters - therefore, sentences belong to those clusters become the reduced search space.

Here, we measure searching time. We run our system with an Intel Xeon 2.20GHz CPU and 196GB memory space.
We pick 20K, 50K and 100K of random sentences from our threat report corpus as for a search space, and randomly pick 5-word, 10-word size query sentences (10 sentences per each and measure their average/minimum/maximum). Then, we measure the results from original search (w/o \textbf{OPT}), search with \textbf{GC} and both-enabled, i.e., \textbf{GC+SC}.

\captionof{table}{\tabular[t]{@{}l@{}} Query Time Measurement \\ (\textbf{a.}:average, \textbf{m.}:min, \textbf{M.}:max) \endtabular}
\label{table:efficiency-test-full}
\begin{minipage}[b]{.5\textwidth}
\begin{tabular}{|c|c|c|c|c|} 
\hline
\multicolumn{2}{|c|}{\multirow{2}{*}{\textbf{Type}}}                                         & \multicolumn{3}{c|}{\textbf{Search Space (\# Sentences)}}  \\ 
\cline{3-5}
\multicolumn{2}{|c|}{}                                                                       & \textbf{20K} & \textbf{50K} & \textbf{100K}                \\ 
\hhline{|=====|}
\multirow{3}{*}{\begin{tabular}[c]{@{}c@{}}\textbf{Baseline}\\(matching)\end{tabular}} & \textbf{a.} & 20s          & 53s          & 1m45s                        \\
                                                                                       & \textbf{m.} & 20s          & 52s          & 1m43s                        \\
                                                                                       & \textbf{M.} & 21s          & 54s          & 1m46s                        \\ 
\hline
\multirow{3}{*}{\begin{tabular}[c]{@{}c@{}}w/o\\\textbf{OPT}\end{tabular}}             & \textbf{a.} & 17m03s       & 43m49s       & 4h16m                        \\
                                                                                       & \textbf{m.} & 16m45s       & 42m57s       & 4h14m                        \\
                                                                                       & \textbf{M.} & 17m15s       & 17m15s       & 4h17m                        \\ 
\hline
\multirow{3}{*}{\begin{tabular}[c]{@{}c@{}}w/\\\textbf{GC}\end{tabular}}               & \textbf{a.} & 28s          & 24s          & 2m44s                        \\
                                                                                       & \textbf{m.} & 11s          & 32s          & 1m04s                        \\
                                                                                       & \textbf{M.} & 44s          & 41s          & 3m55s                        \\ 
\hline
\multirow{3}{*}{\begin{tabular}[c]{@{}c@{}}w/\\\textbf{GC+SC}\end{tabular}}            & \textbf{a.} & 05s          & 14s          & 34s                          \\
                                                                                       & \textbf{m.} & 00s          & 02s          & 05s                          \\
                                                                                       & \textbf{M.} & 14s          & 18s          & 1m19s                        \\
\hline
\end{tabular}\\
\centering
(a) \textbf{query size: 5 words} 
\\
\vspace{2mm}
\end{minipage}
\begin{minipage}[b]{.5\textwidth}
\begin{tabular}{|c|c|c|c|c|} 
\hline
\multicolumn{2}{|c|}{\multirow{2}{*}{\textbf{Type}}}                                         & \multicolumn{3}{c|}{\textbf{Search Space (\# Sentences)}}  \\ 
\cline{3-5}
\multicolumn{2}{|c|}{}                                                                       & \textbf{20K} & \textbf{50K} & \textbf{100K}                \\ 
\hhline{|=====|}
\multirow{3}{*}{\begin{tabular}[c]{@{}c@{}}\textbf{Baseline}\\(matching)\end{tabular}} & \textbf{a.} & 20s          & 53s          & 1m45s                        \\
                                                                                       & \textbf{m.} & 20s          & 53s          & 1m44s                        \\
                                                                                       & \textbf{M.} & 21s          & 54s          & 1m47s                        \\ 
\hline
\multirow{3}{*}{\begin{tabular}[c]{@{}c@{}}w/o\\\textbf{OPT}\end{tabular}}             & \textbf{a.} & 17m58s       & 46m30s       & 4h21m                        \\
                                                                                       & \textbf{m.} & 17m27s       & 45m01s       & 4h18m                        \\
                                                                                       & \textbf{M.} & 2m46s        & 18m46s       & 4h25m                        \\ 
\hline
\multirow{3}{*}{\begin{tabular}[c]{@{}c@{}}w/\\\textbf{GC}\end{tabular}}               & \textbf{a.} & 1m23s        & 4m05s        & 8m03s                        \\
                                                                                       & \textbf{m.} & 53s          & 2m36s        & 5m03s                        \\
                                                                                       & \textbf{M.} & 2m12s        & 2m12s        & 12m41s                       \\ 
\hline
\multirow{3}{*}{\begin{tabular}[c]{@{}c@{}}w/\\\textbf{GC+SC}\end{tabular}}            & \textbf{a.} & 09s          & 28s          & 57s                          \\
                                                                                       & \textbf{m.} & 03s          & 11s          & 25s                          \\
                                                                                       & \textbf{M.} & 23s          & 23s          & 2m21s                        \\
\hline
\end{tabular}\\
\centering
(b) \textbf{query size: 10 words}
\vspace{2mm}
\end{minipage}

Without any optimizations, a query takes from 16m up to 4h based on search space.
However, this can be drastically reduced by our optimization scheme. If we use the \textbf{GC} method, a query time becomes less 10 minutes, and, with \textbf{SC} method enabled, it is expected to be around a few minutes or less than a minute.
Note that 100K of search space is not trivial. We also claim that our system is able to be distributed to multiple machines by dividing the search space.

\begin{algorithm*}
\label{algorithm:isomorphic-discovery-algorithm}
\caption{An algorithm to discover an isomorphic sub-graphs}\label{alg:cap}
\begin{algorithmic}
\State // $V_1$ and $V_2$ have vector nodes (i.e., embedded).
\State $G_1 \gets G(V_1, E_1)$, \, $G_2 \gets G(V_2, E_2)$
\State $S \gets \emptyset$, \, $ \tau > 0$
\State
\For {$v \in V_1$ and $w \in V_2$ s.t. $| v - w | < \tau $}
    \State match $\gets$ false
    \While {$s \in S$}
        \While {$(w', v') \in s$}
            \If {($w \in$ Neighbors($w'$)) $\wedge$ 
                ($v \in$ Neighbors($v'$))}
                \State s.add($(w, v)$) 
                \State match $\gets$ true
            \EndIf
        \EndWhile
    \EndWhile
    \If{match $\neq$ true}
        \State S.add(\{$(w, v)$\})
    \EndIf
\EndFor 
\end{algorithmic}
\end{algorithm*}

\subsection{Dataset Tables}
\label{section:dataset}

This section holds additional tables for our dataset. 

\begin{table}[H]
\centering
\caption{Malware Bahaviors and IoCs Set (\textbf{OI-EVAL-SET-MALWARE}). \textbf{B.} stands for Behaviors}
\vspace{-3mm}
\label{table:dataset-labeled-1}
\begin{tabular}{cccc} 
\hline
\textbf{Malware}        & \textbf{Actor}    & \begin{tabular}[c]{@{}c@{}}\textbf{\# of~}\\\textbf{B.}\end{tabular} & \begin{tabular}[c]{@{}c@{}}\textbf{\# of }\\\textbf{IOCs}\end{tabular}  \\ 
\hline
\textit{Winnti}         & \textit{Axiom}                                                                  & 3                                                                           & 123                                                                     \\
\rowcolor[rgb]{0.902,0.902,0.902} \textit{MessageTap}     & \textit{Axiom}                                                                 & 7                                                                           & 3                                                                       \\
\textit{DTrack}         & \textit{Lazarus}                                                               & 15                                                                          & 25                                                                      \\
\rowcolor[rgb]{0.902,0.902,0.902} \textit{HotCroissant}   & \textit{Lazarus}                                                                & 15                                                                          & 14                                                                      \\
\textit{FrameworkPOS}   & \textit{FIN6}                                                             & 5                                                                           & 25                                                                      \\
\rowcolor[rgb]{0.902,0.902,0.902} \textit{KerrDown}       & \textit{APT32}                                                                & 7                                                                           & 49                                                                      \\
\textit{rDAT}           & \textit{Chrysene}                                                            & 16                                                                          & 6                                                                       \\
\rowcolor[rgb]{0.902,0.902,0.902} \textit{comRAT}         & \textit{Turla}                                                               & 16                                                                          & 16                                                                      \\
\textit{RainyDay}       & \textit{Naikon}                                                                  & 16                                                                          & 10                                                                      \\
\rowcolor[rgb]{0.902,0.902,0.902} \textit{ServHelper}     & \textit{TA505}                                                                & 8                                                                           & 92                                                                      \\ 
\hline
\textbf{\textbf{Total}} & -                                                                                     & 62                                                                          & 363                                                                     \\
\hline
\end{tabular}
\end{table}

\begin{table}[H]
\centering
\caption{Attack Origin Identifying Evaluation Set  (\textbf{OI-EVAL-SET-ACTOR})}
\vspace{-3mm}
\label{table:dataset-labeled-2}
\begin{tabular}{cccc} 
\hline
\textbf{Actor}                                         & \begin{tabular}[c]{@{}c@{}}\textbf{\# of~}\\\textbf{Articles}\end{tabular} & \begin{tabular}[c]{@{}c@{}}\textbf{\# of }\\\textbf{Sentences}\end{tabular} & \begin{tabular}[c]{@{}c@{}}\textbf{\# of }\\\textbf{Words}\end{tabular}  \\ 
\hline
\textit{FIN6}                                                                                 & 17                                                                        & 1,280                                                                       & 21K                                                                   \\
\rowcolor[rgb]{0.902,0.902,0.902} \textit{Leviathan}                                          & 17                                                                        & 1,060                                                                       & 18K                                                                   \\
\textit{Axiom}                                                                            & 21                                                                        & 2,361                                                                       & 38K                                                                   \\
\rowcolor[rgb]{0.902,0.902,0.902} \textit{Stone Panda}                                    & 17                                                                        & 1,333                                                                       & 23K                                                                   \\
\textit{Lazarus}                                                                            & 29                                                                        & 2,338                                                                       & 37K                                                                   \\
\rowcolor[rgb]{0.902,0.902,0.902} \textit{Gorgon}                                           & 22                                                                        & 1,549                                                                       & 23K                                                                   \\
\textit{Turla}                                                                                & 24                                                                        & 1,646                                                                       & 28K                                                                   \\
\rowcolor[rgb]{0.902,0.902,0.902} \textit{TA505}                                             & 27                                                                        & 2,310                                                                       & 34K                                                                   \\
\textit{Chrysene}                                                                         & 21                                                                        & 1,400                                                                       & 25K                                                                   \\
\rowcolor[rgb]{0.902,0.902,0.902} \textit{APT32}                                      & 20                                                                        & 1,430                                                                       & 26K                                                                   \\
\textit{Naikon}                                                                            & 21                                                                        & 1,109                                                                       & 20K                                                                   \\
\rowcolor[rgb]{0.902,0.902,0.902} \textit{C-Major}                                                                     & 22                                                                        & 1,354                                                                       & 23K                                                                   \\ 
\hline
\textbf{Total}                                                                                                          & \textbf{258}                                                                       & \textbf{16K}                                                                      & \textbf{227K}                                                                  \\
\hline
\end{tabular}
\end{table}

\newpage

\subsection{Continued from Motivation}
\label{section:more_on_motivation_examples}

\begin{figure*}
    \centering
    \vspace{3mm}
    {%
    \setlength{\fboxsep}{2pt}%
    \setlength{\fboxrule}{0.5pt}%
    \includegraphics[clip=true,trim=0mm 0mm 0mm 0mm,width=\linewidth]{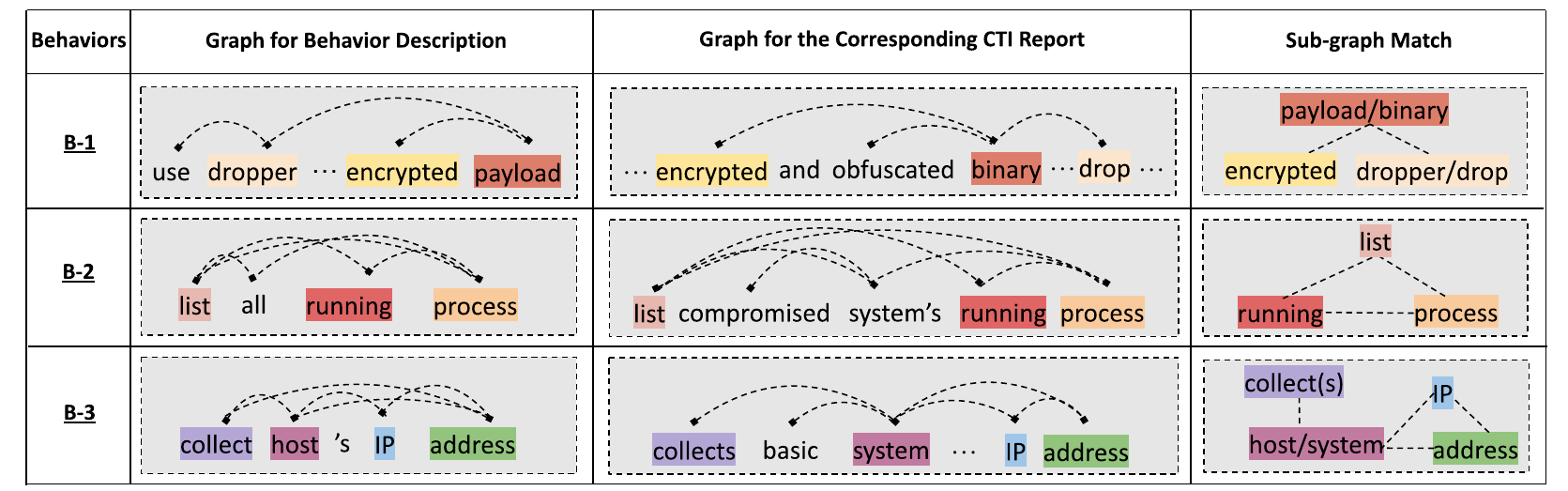}
    }%
    \vspace{-5mm}   
	\caption{Attention graphs for the motivation example}
	\label{fig:how-motivation-example-works}
\end{figure*}

\textbf{How Our System Works.} Figure~\ref{fig:how-motivation-example-works} illustrates how our method works on the motivation example. For each behavior, the second, third, and fourth columns show the attention graph for the IoC description, the graph for the relevant sentence(s) in the corresponding CTI report, and the matched subgraphs. 
For example in {\bf B-1}, our method matches the subgraph including `{\em dropper}', `{\em encrypted}', and `{\em payload'} in the IoC description to that in the report including `{\em drop}', `{\em encrypted}', and `{\em binary}'. Note that in the context of attack forensics, `{\em binary}' is a noun meaning a binary executable file which may be a payload on it own or include a payload. Therefore, our trained language model produces close embeddings for the two. There are similar sub-graph matches for {\bf B-2} and {\bf B-3} as well. $\Box$ \\

\textbf{Google Search.}
Assume the analyst searches \textbf{B-1}, \textbf{B-2} and \textbf{B-3} on Google. 
Most of top search results are not informative. It prioritizes instructional forums due to page ranking and does not bring \textbf{A-1}, \textbf{A-2} and \textbf{A-3}. 

In fact, subtle query differences affect Google search results.  
Our investigation shows that Google fetches the articles (i.e., \textbf{A-1}, \textbf{A-2}, \textbf{A-3}) only if they query ``\textit{encrypted binary that \textbf{eventually} drops}'',  ``\textit{use \textbf{wmi too binary} list running processes}'' and ``\textit{collect \textbf{infected} machine IP addresses}'', respectively which require specific keywords overlapping (words in bold) to retrieve corresponding articles. It also requires tedious scrolling to find them. $\Box$

\end{document}